\documentclass[aps,pre,twocolumn,showpacs,superscriptaddress]{revtex4}
\usepackage{epsfig,amsmath,amssymb,bm,epsf,graphics,graphicx,verbatim,subfigure,color}
\usepackage{slashed}
\newcommand{\be}{\begin{equation}} 
\newcommand{\ee}{\end{equation}}

\def\tone{\theta_1} % The first angular coordinate of a two-rotor system.
\def\ttwo{\theta_2} % The second angular coordinate of a two-rotor system.
\def\tonez{\theta_1(c)} % The first angular coordinate of a two-rotor system along the zero-energy manifold
\def\ttwoz{\theta_2(c)} % The second angular coordinate of a two-rotor system along the zero-energy manifold
\def\sone{\delta_1} % The small deviation from the zero-energy manifold along the first coordinate
\def\stwo{\delta_2} % The small deviation from the zero-energy manifold along the second coordinate
\def\width{w} % width of a soliton
\def\lat{a} % lattice spacing
\def\en{E} % energy
\def\len{\ell} % a length, as in the four-bar linkage
\def\en{U} % The energy of the system

\def\numdim{d} % The number of dimensions, usually two for us
\def\numsites{N} % The number of sites/particles in the system
\def\numbonds{N_B} % The number of bonds in the system
\def\numss{N_{SS}}
\def\coordz{\mathbf{c}} % the zero energy coordinates
\def\vecf{\mathbf{u}} % the vector of finite-energy coordinates
\def\numman{N_0}
\def\coords{\mathbf{r}} % The dN-dimensional vector of the particle coordinates
\def\spring{k} % the spring constant
\def\restlength{\ell} % the rest length of the springs
\def\disp{u} % a small displacement
\def\ext{e} % a small extension
\def\angone{\theta_1} % the first angle coordinate
\def\angtwo{\theta_2} % the second angle coordinate
\def\fe{\mathcal{F}} % Free energy
\def\kb{k_B} % Boltzmann constant
\def\temp{T} % temperature
\def\freq{\omega} % a frequency of a phonon mode
\def\mass{m} % a mass of a particle
\def\dmat{\mathbf{D}} % the dynamical matrix
\def\emat{\mathbf{M}} % an auxiliary matrix used for calculations
\def\spring{k} % The spring system
\def\wave{\mathbf{q}} % the wavenumber
\def\len{\ell} % the length of a bar in, e.g., the four-bar system
\def\width{w} % the width of a soliton
\def\spacing{a} % the lattice spacing
\def\lena{\ell_a} % The side length of the first triangle in the triangle chain
\def\lenb{\ell_b} % The side length of the second triangle in the triangle chain
\def\anga{\psi_a} % The angular width of the first triangle in the triangle chain
\def\angb{\psi_b} % The angular width of the second triangle in the triangle chain
\def\tangone{\theta} % the angle of the first triangle in the cell
\def\tangtwo{\phi} % the angle of the second triangle in the cell

\begin{document}
\title{
Folding mechanisms at finite temperature
}

\author{D. Zeb Rocklin}
\affiliation{Department of Physics, University of Michigan, 450
Church St. Ann Arbor, MI 48109, USA}
\affiliation{Laboratory of Atomic and Solid State Physics, Cornell University, Ithaca, NY 14853}
\affiliation{School of Physics, Georgia Institute of Technology, Atlanta, GA 30332}

\author{Vincenzo Vitelli}
\affiliation{Instituut-Lorentz, Universiteit Leiden, 2300 RA
Leiden, The Netherlands}
\affiliation{The James Franck Institute, The University of Chicago, Chicago, IL 60637}
\affiliation{Department of Physics, The University of Chicago, Chicago, IL 60637}

\author{Xiaoming Mao}
\affiliation{Department of Physics, University of Michigan, 450
Church St. Ann Arbor, MI 48109, USA}

\date{\today}

\begin{abstract}
Folding mechanisms are zero elastic energy motions essential to the deployment of origami, linkages, reconfigurable metamaterials and robotic structures. In this paper, we determine
the fate of folding mechanisms when such structures are miniaturized so that thermal fluctuations cannot be neglected. First, we identify geometric and topological design strategies
aimed at minimizing undesired thermal energy barriers that generically obstruct kinematic mechanisms at the microscale. Our findings are illustrated in the context of a quasi one-dimensional linkage structure 
that harbors a topologically protected mechanism. However, thermal fluctuations can also be exploited to deliberately lock a reconfigurable metamaterial 
into a fully expanded configuration, a process reminiscent of order by disorder transitions in magnetic systems. We demonstrate that this effect leads certain topological mechanical structures to exhibit an abrupt change in the pressure -- a bulk signature of the underlying topological invariant at finite temperature. We conclude with a discussion of anharmonic corrections and potential applications of our work to the the engineering of 
DNA origami devices and molecular robots.  
\end{abstract}

\maketitle

\section{Introduction}

Mechanisms are finite deformations of a structure or mechanical device that cost zero elastic energy. From swinging joints to folding origami and robotic arms, mechanisms are used to perform a variety of functions in natural and technological settings. 
The past decade has witnessed a surge in the efforts to integrate mechanisms in devices at the micro- and nanoscale with potential applications in microrobotics, molecular medicine and nanotechnology. 
The drive towards smaller scales raises an important question concerning the fate of mechanisms in under-constrained structures subject to strong thermal fluctuations. For example, do the folding mechanisms of origami or linkages survive at finite temperature?

An analogous issue arises in the context of polymer physics, where the concept of entropic elasticity was originally formulated to describe floppy systems that acquire rigidities through thermal fluctuations. 
In fact, a flexible polymer can be viewed as a freely jointed chain.  At zero temperature, the chain exhibits exponentially many degenerate ground states, leaving the two ends at arbitrary distance up to its arc-length.  Once thermal fluctuations come into play, the freely jointed chain acquires entropic elasticity, i.e., configurations with two ends of the polymer closer are more favored~\cite{deGennes1979}.
Similar effects are encountered
in cross-linked polymer networks~\cite{deGennes1979,rubinstein2003polymer}, ordered and disordered frames~\cite{Mao2015,Zhang2016,Rubinstein1992,Plischke1998,DennisonMac2013,Mao2013,Mao2013a,Rocklin2014}, graphene kirigami~\cite{blees2015graphene}, self-assembled floppy crystals~\cite{hu2017entropy} and soft spheres below the jamming threshold~\cite{Zhang2009,ikeda2012unified,Degiuli2015}. In all these systems, mechanisms or zero energy modes which cost no elastic energy at $T=0$, become rigid when $T>0$: the free energy of the system is finite even if the potential energy is zero.  

In this paper, we address this problem by considering entropic effects in model mechanical networks which exhibit mechanisms using both analytic theory and Monte Carlo simulations. We pay special attention to a class of zero energy deformations, known as
topological mechanisms, that do not arise from local under-coordination. Instead, their origin can be traced to a topological invariant (akin to the electrostatic polarization) that exists in isostatic structures, i.e., mechanical systems in which case the number of degrees of freedom and constraints are exactly balanced in the bulk~\cite{KaneLub2014,Lubensky2015}. We consider two paradigmatic examples that illustrate the role that geometry and topology play in determining entropic effects in marginally rigid mechanical structures at the micro-scale. 

First, we study a quasi-one-dimensional isostatic structure introduced in Ref.~\cite{KaneLub2014} that possesses a topological polarization and, as a result, supports a zero energy mode of deformation localized at only one edge. When the effect of geometric nonlinearities is taken into account, the localized zero mode becomes a kink whose motion can be initiated only from one edge at a time~\cite{ChenVit2014}. 
Remarkably, the kink can move at zero elastic energy cost without experiencing the typical Peierls Nabarro potential~\cite{braun2013frenkel} arising from the periodicity of the underlying lattice.
However, for thermal systems entropic forces can nevertheless  restore rigidity  (i.e., via a thermally induced Peierls Nabarro potential), potentially hindering the propagation of the kink. In this work, we show that entropic forces on kinks in this topological chain are exponentially suppressed:
the mechanism can survive large thermal fluctuations when miniaturized, provided that the geometry of the unit cell is suitably designed.

By contrast, we study a class of two-dimensional lattices in which thermal fluctuations play a dominant role driving the system to a fully-expanded configuration. If anharmonic terms of fluctuations are ignored, such lattices can exhibit a discontinuity in pressure
that is a clear bulk signature of their topological polarization.
At the transition, fluctuations become large and geometrical (not material) anharmonicity regularizes the apparent discontinuity. Even at zero temperature, these topological mechanical lattices possess a number of striking features, including directional mechanical response~\cite{rocklin2017directional} and the ability to incorporate and exploit
rotational degrees of freedom~\cite{meeussen2016geared}, defects~\cite{paulose2015topological} and
bulk topological modes~\cite{rocklin2016mechanical}.

The rest of this paper is organized as follows. In Section II, we develop the theory of generic frames at finite temperature and illustrate it with the paradigmatic example of a four-bar linkage in Section III. In Section IV, we study kink propagation at finite temperatures in the one-dimensional topological chain, relegating most of the relevant mathematical derivations to Appendices A and B. In Section V, we consider two-dimensional lattices.
In Section VI, we discuss the effect of anharmonic terms in the Hamiltonian, which can regularize singularities in the free energy at finite $T$.  
In Section VII, we propose experimental realizations involving DNA tile structures composed of stiff DNA beams joined together by relatively flexible hinges, of which we build a macroscopic prototype.

\section{Generic frames at finite temperature}
\label{sec:generic}

Consider a number $\numsites$ of particles in $\numdim$ dimensions whose positions $\{\coords_i \}$ can then be described by the $ \numsites\numdim$-dimensional vector $\coords$. These particles are subject to $\numbonds$ pairwise central-force interactions of the form

\begin{align}
V( |\coords_i - \coords_j|) = \frac{1}{2} \spring \left(  |\coords_i - \coords_j| - \restlength_{ij} \right)^2,
\end{align}

\noindent representing spring-like bonds of strength $\spring$ connecting pairs of particles.  ``Mechanisms'' in this system correspond to ways in which particles move by an arbitrarily large amount relative to one another without changing the length of any bond.  For any given configuration of the system, the mechanisms present are revealed by 
a linear analysis of zero modes.  If a system possesses mechanisms, such linear analysis can be done at each configuration following mechanisms of the system, and we refer to the space of these configurations as the ``zero-energy manifold'' of the system.  

Interestingly, the dimension of the zero-energy manifold may be different at different configurations.  The reason is that a previously independent bond can become redundant in certain configurations.  To characterize this, we introduce a ``nonlinear'' version of the counting of zero modes

\begin{align}
\numman(\coordz) = \numsites\numdim  - \numbonds - \numss(\coordz),
\label{eq:indexeqn}
\end{align}

\noindent where $\coordz$ are the coordinates along the zero-energy manifold.  
Previously, such counting has been used to give the number of zero modes present in a linearized system~\cite{Maxwell1864,Calladine1978}, and has been shown to serve as the index theorem in the topological mechanics of Maxwell lattices~\cite{KaneLub2014,Lubensky2015}.   
All bonds are assumed to be at their rest lengths in the configurations where this counting is applied (e.g., it does not apply to the ``tensegrity'' structures as introduced in Ref.~\cite{Calladine1978}).   
The last term $\numss(\coordz)$ is the number of states of self stress at coordinate $\coordz$.  States of self stress are ways of distributing forces on bonds leaving no net force on any particle, and their relation to redundant bonds was rigorously proved in Ref.~\cite{Calladine1978}. 
This equation can be understood through the following simple analysis: each bond of the system either is an independent constraint, and reduces $\numman$ by one, or it can be redundant, and contribute a state of self stress to the system.  

Different from previous studies, here we consider how this equation evolves in the zero-energy manifold, whose dimension is 
\begin{align}
	d_m(\coordz)  = \numman(\coordz) - \frac{\numdim(\numdim+1)}{2},
\end{align}
where we by convention remove the trivial zero modes corresponding to the rigid translation and rotation of the whole system and leave only mechanisms.  It is straightforward to conclude that when states of self stress arise as a result of certain bonds becoming redundant, the dimension of the zero-energy manifold increases,
a process we will illustrate in Sec.~\ref{sec:fourbar} with the \emph{four-bar system}, also known as a Bennett linkage. Marras et al. have realized such a mechanism in DNA origami and actuated it via the release of additional DNA strands~\cite{Marras2015}.

At zero temperature, infinitesimal external force can deform a system along mechanisms without any energetic cost.
However, at finite temperature thermal fluctuations excite finite-energy modes. Over time scales longer than the thermalization time of the system, behavior of the system is governed by its free energy $\fe=\en-TS$.  Along coordinates of a mechanism, although the elastic energy $U=0$ everywhere, entropy $S$ can vary significantly.  To characterize this, we adopt the canonical ensemble of the elastic system to calculate its free energy for any point on the manifold

\begin{align}
\label{eq:fe}
\fe\left(\coordz\right) &= - \kb \temp \log \int d \vecf \exp \left[\vecf^T \cdot \dmat\left(\coordz\right)\cdot \vecf/\kb \temp \right] \nonumber\\
&=
\kb \temp \sum_i \log \frac{\mass \freq_i\left(\coordz\right)^2}{2 \pi \kb \temp},
\end{align}

\noindent where $\dmat\left(\coordz\right)$ is the dynamical matrix, which varies nonlinearly on the zero-energy manifold. The sum is over all normal modes $\{\vecf_i\}$ with frequencies  $\{ \freq_i \}$ for small vibrations, $\kb$ is Boltzmann's constant and $\temp$ denotes the temperature. .
When a system has zero modes, the corresponding frequencies $\freq_i$ of these modes vanish, leading to a divergent contribution to $\fe$.  However, the same divergent terms are present for all configurations along the same manifold.  Since we are interested in the change of the free energy $\Delta \fe$ between different configurations in the same manifold, the divergent terms cancel out, and $\Delta \fe$ is controlled by the change in $\freq_i$ of the modes in the space orthogonal to the zero-energy manifold.  

Three observations immediately follow this formulation.  First, $\fe$ is lower at configurations in which the frequencies of nonzero modes are relatively lower, and $\fe$ reaches singularities when additional floppy mode - self stress (FM-SS) pairs arise where the zero-energy manifold self-intersects.  This shares interesting similarities with the ``order-from-disorder'' effect in frustrated spins~\cite{Shokef2011,Villain1980,Shender1982,Henley1987,Henley1989,Chubukov1992,Reimers1993,Bergman2007}.  The emergence of extra FM-SS pairs leads to logarithmic divergence terms in the free energy, which is an artifact of the quadratic theory in Eq.~\eqref{eq:fe}.  In Sec.~VI, we discuss how this singularity is regularized by geometric nonlinearities in the system.
Second, when two parts of the zero-energy manifold are very close but not crossing, thermal fluctuations can allow the system to jump over the region of nonzero energy, leading to ``tunneling'' between disconnected parts of the zero energy manifold.  
Third, in the case when all spring constants take the same value $k$, a factor of $k$ can be taken out from all  $\freq_i$, leaving a constant $\kb \temp  \numsites \numdim \log k $ in the free energy.  Thus the difference $\Delta \fe$ is a pure geometrical quantity times $\kb T$ which is independent of the scale of elastic energy in this case.

The consequence of $\fe$ varying along the zero energy manifold (where $U=0$) is that
the mechanism is no longer a deformation which requires no external work to operate.
  A system with no external drive spends more time in a configuration $\coordz_1$ with low free energy than a second, $\coordz_2$ with high free energy by a factor
 $\exp \left[\left(\fe\left(\coordz_1\right)-\fe\left(\coordz_2\right)\right)/k_B T\right]$. 
	In order to drive the system through the mechanism, in general, one has to do work, unless $\fe\left(\coordz\right)$ is very flat along the mechanism.  Such an interesting example will be discussed in Sec.~\ref{sec:chain}.

\section{Four-bar linkage}
\label{sec:fourbar}
\begin{figure}
\includegraphics[width=0.49\textwidth]{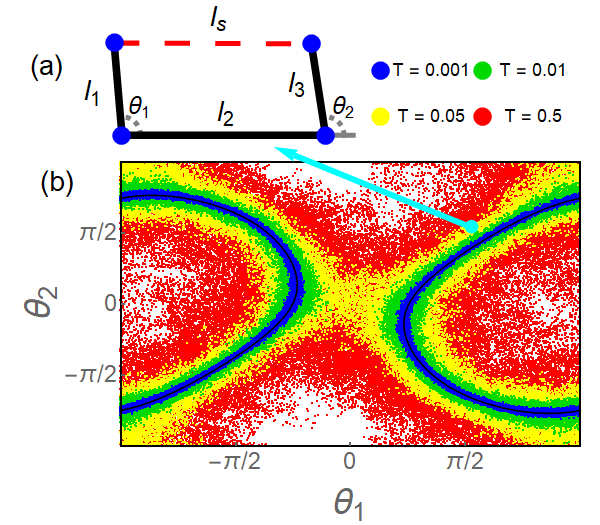}
\caption{
(a) A four-bar linkage, with three rigid bars of lengths $\ell_1,\ell_2,\ell_3$ and the fourth a Hookean spring whose dynamical length $\ell_s$ may differ from its equilibrium value $\ell_4$. The configuration, given by angles $\theta_1,\theta_2$ of the first and third bars relative to the second, compresses/stretches the spring $\ell_s$.
(b) The linkage has a mechanism in which bars may rotate without compressing/stretching, leading to the one-dimensional zero-energy manifold (black curves). At finite temperature, the linkage distorts the spring and assumes finite-energy configurations. In Monte Carlo simulations of increasing temperature, as one moves away from the zero-energy manifold the samples points shown become increasingly nonlinear (green) begin to tunnel from one zero-energy configuration to another (yellow) and eventually rotate increasingly independently (red). Note that even at low temperatures (blue) the \emph{ratio} of fluctuations around different zero-energy configurations remains finite.
}
\label{fig:simresults}
\end{figure}

We first use a simple example to illustrate the zero-energy manifold and the variation of the free energy in this manifold.
Consider the four-bar linkage shown in Fig.~\ref{fig:simresults}(a), where bars are connected by free hinges.
Provided that no bond length $\ell_i$ is as great as the sum of the other three, Eq.~(\ref{eq:indexeqn}) indicates that such a linkage will have at least one floppy mode in addition to two rigid translations and a rotation. We have then a one-dimensional zero-energy manifold
($d_m(\coordz)$=1 except for configurations where FM-SS pairs arise). For conceptual simplicity, we restrict three bonds to be of fixed length ($\spring \to\infty$) such that the energy contributions come entirely from extensions of the fourth bond and the space of all configurations is two-dimensional and parametrized by the angles $\angone, \angtwo$, as used for individual segments of the topological rotor chain~\cite{KaneLub2014, ChenVit2014}.

\begin{figure*}
\includegraphics[width=0.98\textwidth]{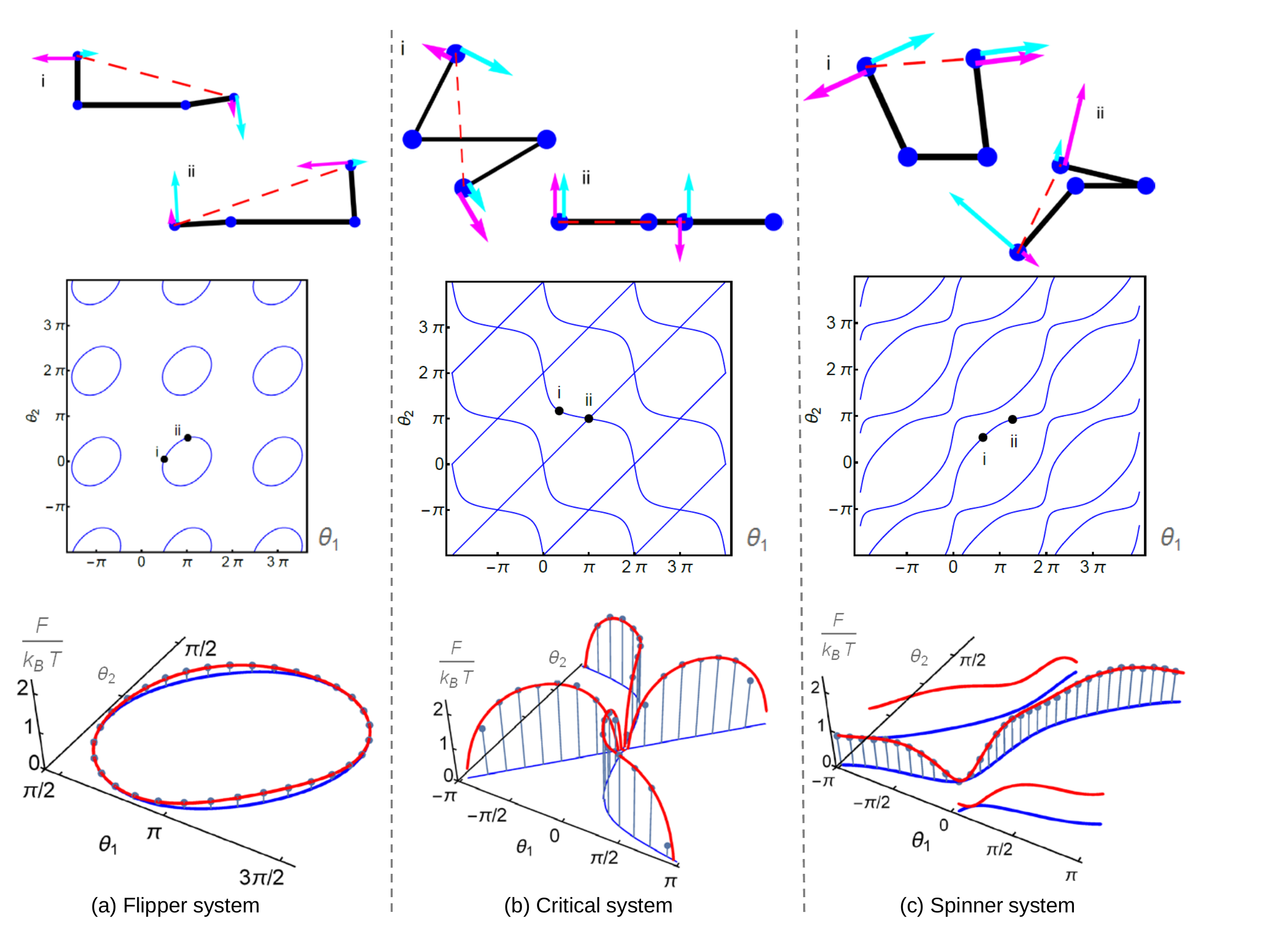}
\caption{
Four-bar linkages (diagrams, top) subject to thermal fluctuations that rotate the bars. Every configuration has a zero-energy deformation (denoted by cyan arrows) that doesn't stretch the spring (red dashed line) and a generally-finite energy deformation that does (magenta arrows).
(a) A ``flipper'' linkage doesn't allow full rotation of the bars connected to the spring, following the closed (blue) path. Fluctuations are largest when the bars are almost parallel, favoring the configuration of the right diagram over the left with a lower free energy (red line via theory, points via Monte Carlo simulation).
(b) A critical linkage, with a critical configuration (right diagram) in which bars are exactly parallel and there are two zero modes, along the two paths of the zero-energy manifold.
(c) A ``spinner'' linkage allows the bars connected to the spring to fully rotate. As with the other linkages, transverse fluctuations favor some configurations over others.
}
\label{fig:fourbar}
\end{figure*}

Generically, the zero energy manifold is not connected. In particular, in order for $\angone = 0, \pi$ to be achievable without stretching any bond the triangle inequality dictates respectively

\begin{subequations}
\begin{align}
\label{eq:teq}
|\len_1-\len_2| &\ge |\len_3-\len_4|, \\
\len_1+\len_2 &\le \len_3 + \len_4.
\end{align}
\end{subequations}

\noindent These conditions are not always satisfied. Fig.~\ref{fig:fourbar}(a) shows an example in which neither angle may undergo a full revolution. For both of these conditions to be satisfied for $\angone$ would require

\begin{align}
\label{eq:teq2}
\textrm{min}(\len_1,\len_2) \le \textrm{min}(\len_3,\len_4).
\end{align}

\noindent Even when this condition and the analogous one {for $\angtwo$ are satisfied, they cannot guarantee} 
free rotation along both of the other two pivots, so that the zero-energy manifold shown in Fig.~\ref{fig:fourbar}(c) is still generally disconnected. The only exception is when one of the expressions is satisfied with equality, in which the zero-energy manifold intersects itself.
The other two  situations, in which a particular bar either flips back and forth in a finite range or spins continuously from $0$ to $2\pi$ are referred to as the ``flipper'' and ``spinner'' states respectively in Ref.~\cite{ChenVit2014}.

In contrast, when $\len_1 = \len_3, \len_2=\len_4$, all such inequalities are satisfied and the zero-energy manifold becomes connected, as shown in Fig.~\ref{fig:fourbar}(b). Such a system contains two critical points $\angone=\angtwo=0$ and $\angone=\angtwo=\pi$, collinear configurations in which FM-SS pairs arise. At these critical points, the one-dimensional zero-energy manifold becomes two-dimensional, with the two paths corresponding to two different nonlinear buckling modes. More generally, when only one of the expressions of Eq.~(\ref{eq:teq}) is satisfied with equality there exists one critical point, and we refer to any such system as a ``critical system''. The critical point is topological in that passing through it alters whether one of the angles increases (or decreases) by one revolution as the linkage undergoes a circuit along the zero-energy manifold~\cite{ChenVit2014}.

Next, in order to account for small thermal fluctuations we consider a point $(\tone,\ttwo)=(\tonez,\ttwoz)+(\sone,\stwo)$ a small distance from the zero-energy manifold, leading to an energy functional that is nonlinear in the position along the manifold but quadratic in the small fluctuations therefrom:
\begin{align}
\en =& \frac{\spring}{2} \left(\len_s(\tone,\ttwo)- \len_4\right)^2;
 \\
\len_s^2(\tone,\ttwo) =& 4 \sin^2\left(\frac{\tone-\ttwo}{2}\right)\len_1^2\,
+ \nonumber
\\
&2 \len_1 \len_2 \left(\cos \ttwo - \cos \tone\right) + \len_2^2;
\end{align}
leading to
\begin{align}
\en \approx &
\frac{k}{2} \left( \frac{\len_1}{\len_4}\right)^2
\big[
\sone
\left(
\len_2 \sin \tonez + \len_1 \sin(\tonez - \ttwoz)
\right)
\nonumber
\\
&-
\stwo
\left(
\len_2 \sin \ttwoz + \len_1 \sin(\tonez - \ttwoz)
\right)
\big]^2,
\end{align}

\noindent
{where for convenience we have specialized to the symmetric systems for which $ \len_1= \len_3$. }
This small-fluctuation approximation  is generally warranted when the temperature is low compared to the characteristic spring energy scale $\spring \len^2$, though geometric alignment of bonds can lead to larger fluctuations.
Of our two normal frequencies of small oscillations $(\sone,\stwo)$, the one for the mode along the zero-energy manifold is always zero, leading to a divergent but constant contribution to the free energy as in Eq.~(\ref{eq:fe}). The remaining mode, however, lowers the free energy at points on the zero-energy manifold in which transverse fluctuations are large, so that the system will spend more time in those configurations. The zero-energy manifold is a valley in the energy landscape whose depth is constant but whose width varies, influencing the equilibrium configurations of the linkage {(Fig.~\ref{fig:simresults}b).

The variation of $\fe$ along the zero-energy manifold reaches an extreme in the critical systems [Fig.~\ref{fig:fourbar}(b)], because at self-intersecting points both of the normal modes have $\freq_i=0$, {leading to points with logarithmic divergences in the free energy,
\begin{align}\label{EQ:FELOG}
	\fe \sim k_B T \log |\theta_1(c) - \theta_1(c_0)| ,
\end{align}
where $c_0$ denotes the self-intersecting point on the zero-energy manifold.  As we discuss below in Sec.~VI, very close to the crossing point, terms anharmonic in $(\sone,\stwo)$ becomes important.  Thus, this logarithmic divergence from quadratic theory only characterizes the free energy variation when the system is not too close to the crossing point.}

In addition, very close to the critical system, because two sections of the zero-energy manifold are very close, when the barrier between them is comparable to thermal fluctuations, the system can tunnel between these sections of the zero-energy manifold (Fig.~\ref{fig:simresults}b).  Thus, to fully characterize finite temperature behavior of close-to-critical systems, potential energy terms beyond quadratic order should be considered.  This can be done either via Monte Carlo simulation (Fig.~\ref{fig:simresults}) or by analytically considering the renormalization of the mode rigidities as done in Ref.~\cite{Mao2015}.  We discuss such regularization in Sec.~\ref{sec:regularization}.

To summarize the four-bar linkage analysis, we find that in general mechanisms are lifted to finite free energy variation when $T>0$ so they require finite drive to operate, in contrast to the $T=0$ case where infinitesimal force (ignoring friction) can drive the system through the whole mechanism while the elastic energy remains zero.   Our analysis also shows that in order to minimize the free energy difference to get a smooth mechanism, one should choose a parameter set where the linkage is far from the critical system and the finite mode frequency does not exhibit large changes.

\section{1D chains}
\label{sec:chain}

We now consider a class of one-dimensional mechanical lattices consisting of individual elements that, like the four-bar linkages of the previous section, are capable of undergoing nonlinear, zero-energy deformations. We consider primarily the rotor chain introduced in~\cite{KaneLub2014}, which
consists of $N$ rotors and $N-1$ springs, leading to one floppy mode.  It was shown that this floppy mode localizes to either the left or right end of the chain, controlled by the topological polarization in the bulk of the chain~\cite{KaneLub2014}.  Following the zero energy manifold of the chain to nonlinear order, the edge floppy mode turns into a domain wall kink (what could loosely be called a soliton, though generally without properties such as integrability) traveling through the bulk, and flips the topological polarization of the domain it passed through~\cite{ChenVit2014}.
This non-linear wave, easily realized on the macroscale, can convey force and information. In Sec.~\ref{sec:implementation} we present a similar structure, which we call the triangle chain, that could be realized via, e.g., DNA origami. When translated to the microscale, thermal fluctuations become significant, raising the possibility that the mechanism will be driven into undesired configurations or will acquire significant resistance against advancing along the chain, analogous to the properties of the four-bar linkages of the previous section. We now identify key design principles to avoid this fate.

We consider a class of systems whose configurations, at any point $x$ are described by a continuous coordinate $\theta(x)$
whose evolution across space is given in the zero-energy manifold by some function $s(\theta)$ such that $d \theta/d x = s(\theta)$, which leads to a one-dimensional zero-energy manifold determined by the value of the coordinate at a single point. This is enforced by the energy functional

\begin{align}
\label{eq:chainenergy}
\en = \frac{\spring}{2}\int d x \, \left[ \frac{d{\theta}}{d x} - s(\theta)\right]^2.
\end{align}

\noindent Considering only systems that are symmetric under reflection $\theta \rightarrow - \theta, x \rightarrow -x$ but that do not admit the symmetric solution $\theta(x)=0$, leads to even slope functions of the form $s(\theta) = c\left(\bar{\theta}^2-\theta^2\right)+ O(\theta^4)$, ensuring that the zero-energy configuration  can vary smoothly between two equilibrium configurations $\pm \bar{\theta}$. Indeed, this form of slope function is the only one symmetric under $\theta \to -\theta, x \to -x$, possessing nontrivial uniform solutions, and not containing higher-order terms. As such, it applies not only to our particular systems but to a broad class of symmetric mechanisms in one-dimensional structures undergoing small deformations.

As shown in Ref.~\cite{ChenVit2014}, this leads to the zero-energy kink (or domain wall) profile characterized by the amplitude $\bar{\theta}$, width $w = (\bar{\theta}c)^{-1}$ and center $x_c$:

\begin{align}
\label{eq:soliton}
\theta(x) = \bar{\theta} \tanh \left[(x-x_c)/w\right].
\end{align}

\noindent	Note that the anti-kink solution (corresponding to the substitution $\bar{\theta} \rightarrow -\bar{\theta}$ in Eq. (\ref{eq:soliton})) does not belong to the zero energy manifold $U=0$ defined by Eq. ({\ref{eq:chainenergy}}). 

For discrete, lattice systems whose coordinate remains small enough that the higher-order terms may be neglected, the energy functional takes the form

\begin{align}
\label{eq:chainenergydiscrete}
\en = \frac{u_0 a^2}{2}\sum_j \, 
\left[ 
\frac{\theta_{j+1} -\theta_j}{\lat} - 
c\left[\bar{\theta}^2-\left(
\frac{
\theta_j+\theta_{j+1}
}{2}
\right)\right]^2
\right]^2,
\end{align}

\noindent where $\lat$ is the lattice spacing and $u_0$ has units of energy. Describing the configurations of Eq.~(\ref{eq:chainenergydiscrete}) in terms of zero-energy kink configurations $f_j$ and small finite-energy excitations $u_j$ so that $\theta_j=f_j+u_j$, the energy to $O(u^2)$ is

\begin{align}
\label{eq:chainenergydiscretetwo}
U &= \frac{u_0}{2} \sum_j \left( a_j u_j - b_j u_{j+1}\right)^2, \textrm{with} \nonumber \\
a_j(b_j) &= 1 \pm \lat c \left( f_{j+1} +f_j\right)/2.
\end{align}

\noindent As derived in Appendix~\ref{APP:FE}, the exact free energy for such systems is

\begin{align}
\label{eq:prodsum}
\fe = \frac{T}{2} \log \left[ \sum_{j'} \left( \prod_{j<j'} a_j^2 \right) \left( \prod_{j \ge j'} b_j^2 \right) \right].
\end{align}

 A simpler expression, though, is the form derived from making a Peierls-Nabarro-type approximation~\cite{braun2013frenkel}, in which each spring is treated as a component of a \emph{separate} four-bar linkage of the type described in Sec.~\ref{sec:fourbar}.  Analyzing normal modes of each four-bar linkage in the chain, we find the stiffness of the nonzero mode of the linkage to be proportional to $\sqrt{a_j^2 + b_j^2}$.
%, giving it a single degree of freedom with effective spring constant proportional to $\sqrt{a_j^2 + b_j^2}$ 
This leads, for weak solitons, to an approximation for the free energy

\begin{align}
\fe &\approx \frac{T}{2} \sum_j \log \left[ 1 + \frac{\lat^2 c^2}{2}\left(f_{j+1} + f_j \right)^2 \right]\nonumber \\ &\approx \frac{T}{2} \lat^2 c^2 \sum_j \left(\frac{f_{j+1}+f_j}{2}\right)^2.
\end{align}

\begin{figure}
\centering
\includegraphics[width=.45\textwidth]{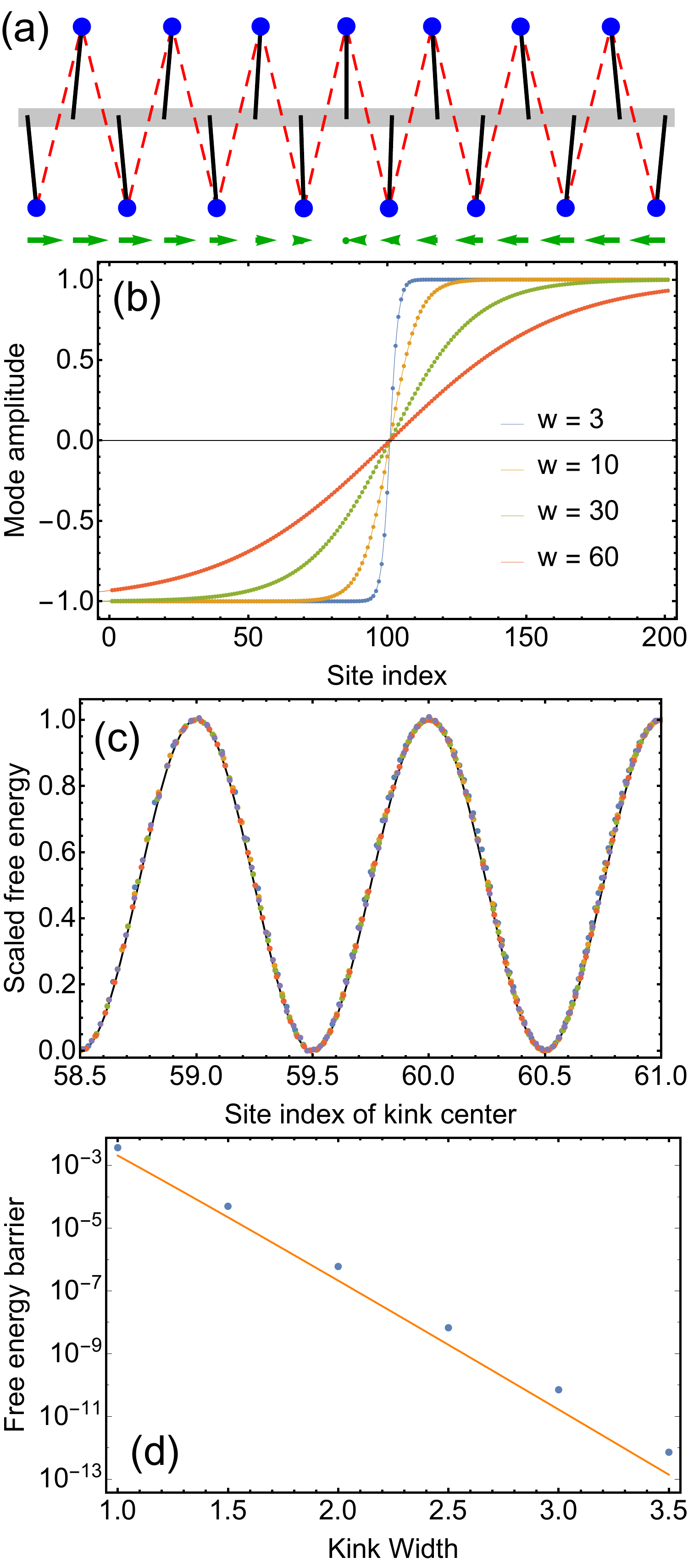}
\caption{
(a) A chain of rotors has a zero-energy soliton-like kink that propagates freely at zero temperature.
(b) The kink of the discrete $\phi^4$ system varies widely in width, conforming closely to the continuum theory when the kink is not too narrow.
(c) Thermal fluctuations favor certain configurations in the lattice over others, resulting in an effective sinusoidal potential which favors soliton configurations centered \emph{between} rather than \emph{on} lattice sites.
(d) For kinks whose width $\width$ is comparable to or greater than the lattice spacing $\lat$, the free energy barrier is exponentially suppressed, permitting free motion of the soliton.
}
\label{fig:onedeechain}
\end{figure}

\noindent 
%This energy functional leads, via Eq~(\ref{eq:fe}), to a free energy as a function of the position of the center of mass, $\mathcal{F}(x_{cm})$. 
This free energy is a function of the center of mass of the soliton.  
In the lattice case, this free energy is not constant but periodic with the spatial period $a$ of the lattice. We
 may replace the sum over sites $j$ with an integral over positions $x$ by including the Dirac comb, $\sum_j \delta(x-j \spacing)=\sum_n
\exp(2 \pi i n x/\spacing)/\spacing$. For wide solitons
one can make the approximation $(f_{j+1}+f_j)/2\simeq \theta(x_j)$ and use Eq.~(\ref{eq:soliton}), and find that
% such that the form Eq.~(\ref{eq:soliton}) is still valid and the free energy density is small, 
the free energy of the soliton relative to the soliton-free chain is

\begin{align}
\label{eq:chainfe}
\fe(x_c) = - \temp \frac{ \lat}{\width} - 2 \pi^2 \temp  \sum_{n=1}^\infty \frac{n \cos ( 2 \pi n x_c/\spacing)}{\sinh(n \pi^2 \width/\spacing)}.
\end{align}

\noindent The first, negative term indicates that the soliton lowers the free energy of the chain, discouraging the soliton from escaping to the boundaries. The leading non-constant term is sinusoidal with the period of the lattice and a height exponentially small in the soliton width, leading to a free energy barrier 
\begin{align}
	\Delta \fe \sim 8  \pi^2 \temp \exp(-\pi^2 w /a) . 
\end{align}
As shown in Fig.~\ref{fig:onedeechain}, this approximation qualitatively matches the behavior of the numerical result derived via Eq.~(\ref{eq:prodsum}) even for minute free energy barriers. As shown in Appendix~\ref{APP:EXPO}, self-averaging of phonon modes leads generally to these exponentially small barriers for smooth kink profiles, ensuring smooth propagation of the wave even under thermal fluctuations.

In order to operate smoothly, the kink must also have a sufficiently high amplitude to prevent the generation of defects via thermal fluctuations. Fortunately, Eq.~(\ref{eq:chainfe}) shows that the free energy barriers are insensitive to this amplitude. Thus, we find our design principle: kink modes will operate smoothly under thermal fluctuations provided that their amplitudes are large enough to suppress tunneling fluctuations and they are somewhat wider than the underlying lattice.

This behavior is depicted in Fig.~\ref{fig:onedeechain}. The kink shown in (a), despite having a width only slightly greater than the lattice spacing, experiences miniscule free energy barriers four orders of magnitude smaller than the thermal energy. As shown in (b), these barriers fit well to a sine wave with the periodicity of the lattice, with
further corrections exponentially smaller than the leading correction, a phenomenon characteristic of Peierls-Nabarro potentials.
Unlike the standard PN potential the amplitude of the barriers is set by the thermal energy scale. This free energy barrier shrinks rapidly as the kink width is increased, indicating that the mechanism can survive quite naturally in thermal systems. In the next section, we consider a class of systems for which this proves not to be the case.

\section{2D lattices}
\label{sec:twodee}

In this section we consider effects of thermal fluctuations on two-dimensional (2D) lattices with mechanisms.  We focus on Maxwell lattices, which are lattices with point-like particles connected by central-force springs, with mean coordination number $\langle z\rangle =2d$, leaving the lattice at the verge of mechanical instability~\cite{Lubensky2015}.  Maxwell lattices have been used to characterize a broad range of systems near the onset of rigidity and provided useful insight on rigidity transitions~\cite{Souslov2009,MaoLub2010,Mao2011a,Ellenbroek2011,
Broedersz2011,MaoLub2013a,MaoLub2011b,Zhang2015a,Zhou2017,Zhang2018}.  

Maxwell lattices are shown to always exhibit $d(d-1)/2$ homogeneous mechanisms which change the lattice geometry (either corresponding to macroscopic strains and called ``Guest-Hutchinson'' (GH) modes or corresponding to homogeneous modes which involve pure rotations of lattice components and vanishing strain)~\cite{GuestHut2003,Lubensky2015}.  Examples of such homogeneous mechanisms are shown in Fig.~\ref{fig:twodee}.

\begin{figure}
\centering
\includegraphics[width=.49\textwidth]{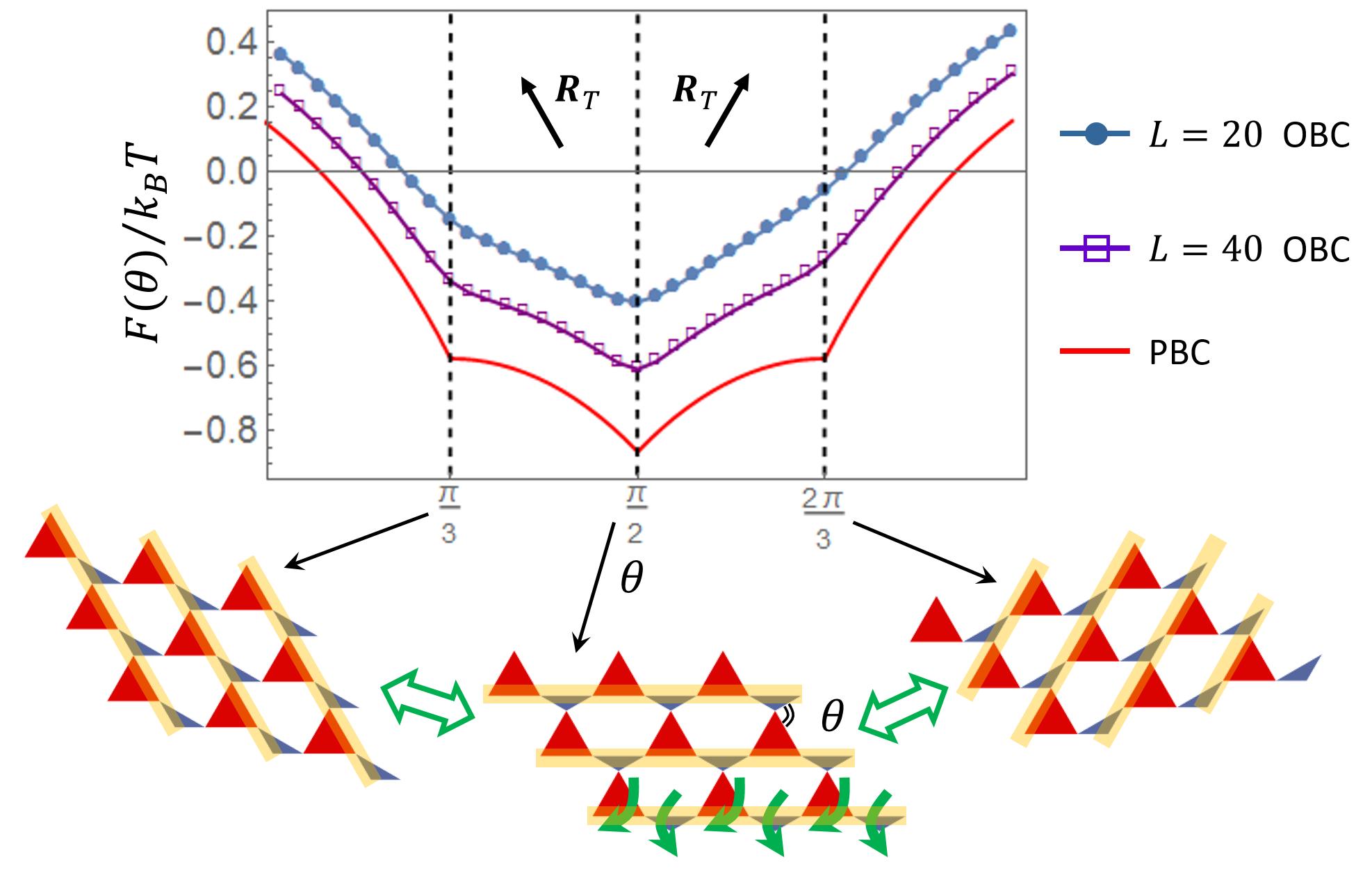}
\caption{
Free energy per unit cell $\fe(\theta)$ along the zero energy manifold (GH mode, labeled by bond angle $\theta$) of a 2D topological kagome lattice.  The three curves show $\fe(\theta)$ for a $20\times 20$, $40\times 40$ with open boundary conditions (OBC) [following Eq.~\eqref{eq:fe}], and the periodic boundary conditions (PBC) [numerical integral following Eq.~\eqref{EQ:FtwoD}].  Below the plot we show geometries of the lattice at the three critical states where the topological polarization (shown in the plot) of the lattice changes.  These critical states are characterized by straight lines of bonds (shown in yellow stripes) and floppy modes in the bulk (green arrows in the middle figure).  Unit cell geometry of the lattice is chosen such as the red (pointing up) triangles have side lengths $(1,1,1)$ and the blue (pointing down) triangles have side length $(1,1/\sqrt{3},1/\sqrt{3})$.  As the result the three critical states have twisting angles $\pi/3, \pi/2, 2\pi/3$.  
%Kagome_T_finite_size.nb
}
\label{fig:twodee}
\end{figure}

In particular, we consider topological kagome lattices, which are found to exhibit edge floppy modes controlled by the topological polarization of the lattice.  The topological polarization of these 2D lattices belongs to the same class as the topological polarization of the 1D chain we discussed in Sec.~\ref{sec:chain}, 
although here there are two winding numbers, defined for the two primitive lattice directions of the 2D lattice, and thus the topological polarization is a vector, $\mathbf{R}_T$.  
   It is shown that when the topological kagome lattice follows the soft strain of the GH mode (the lattice is 2D and thus has only one GH mode, for which we use the bond angle $\theta$ as the coordinate $\coordz$), the lattice goes through critical states in which the topological polarization $\mathbf{R}_T$ switches (Fig.~\ref{fig:twodee})~\cite{rocklin2017transformable}.  This shares an interesting similarity with the kink in the 1D chain we discuss in Sec.~\ref{sec:chain}, which also changes the topological polarization of the whole system.  There, the kink can remain a smooth mechanism even at $T>0$;  what happens to the GH modes at finite $T$?

Because the GH modes preserve periodicity, it is now convenient to calculate the free energy over the Brillouin zone in terms of the dynamical matrix $\dmat \left(\theta,\wave\right)$ depending on both the bond angle $\theta$ along the GH mode (as the zero energy manifold coordinate) and the wavevector $\wave$:

\begin{align}\label{EQ:FtwoD}
\fe\left(\theta\right) = \frac{\kb \temp}{2} \int_{\textrm{BZ}} d^2 \wave \log \det \dmat \left(\theta,\wave\right).
\end{align}

Similar to the cases of the four-bar linkage and the 1D topological chain, this free energy is lower when phonon mode frequencies are lower as the lattice moves along the mechanism coordinate $\theta$, and exhibits singularities when extra FM-SS pairs emerge.  For the case of the topological kagome lattice, interestingly, when the lattice passes through the critical state where topological polarization of the lattice changes, bonds form straight lines, giving rise to extra FM-SS pairs on lines in $\wave$ space.  Upon integration over the whole Brillouin zone, this leads to singularities in $\fe$ as the lattice passes through the critical state, as shown in Fig.~\ref{fig:twodee}.

{In particular, these free energy singularities are characterized by the form
\begin{align}\label{EQ:FE2DASYM}
	\fe\left(\theta\right) \sim \kb \temp |\theta - \theta_c| +\fe_{\rm{n.s.}}\left(\theta\right) ,
\end{align}
where $\theta_c$ is the GH mode coordinate of the critical lattice where extra FM-SS pairs emerge (general topological kagome lattices have 3 critical states as shown Fig.~\ref{fig:twodee}), and the second term denotes non-singular terms of the free energy.  This form can be obtained via asymptotic analysis of the phonon spectrum of the deformed kagome lattice, which exhibits FM lines of frequency $|\theta - \theta_c|$ in the Brillouin zone~\cite{SunLub2012,KaneLub2014}.  Note here the singularity is in the form of $|\theta - \theta_c|$, rather than $\log |\theta - \theta_c|$, because of the integral in the 2D Brillouin zone.  
More discussions of the free energy singularities when anharmonic corrections are considered is included in Sec.~VI.
}

It is worth pointing out that although this calculation of free energy assumes periodic boundary condition in order to integrate in momentum space, the change of the free energy along the mechanism still provides useful information when the lattice is under open boundary conditions, which is the case in experiments.  Under open boundary condition, the total number of floppy modes never changes as the lattice follows the mechanism.  However, the rest of the phonon modes at nonzero frequencies, which are not sensitive to boundary conditions, are lowered when the lattice is at its critical state, and they contribute a significant change in the free energy, although the singularity is smoothed out.  As we show in Fig.~\ref{fig:twodee}, the free energy of finite lattices under open boundary conditions approaches the periodic boundary condition result as the lattice size increases.

The change in free energy as the 2D topological lattice deforms along its mechanism governs the macroscopic mechanical properties of the lattice at finite temperature.  The free energy we calculate is a function of strain, and its differential can be written as $d\fe = -SdT + \sigma d\epsilon$ in its most general form, where $\sigma$ and $\epsilon$ stand for stress and strain.  For Maxwell lattices we are particularly interested in deformations along the GH mode (zero energy manifold) $\theta$, so we specialize to free energy  $d\fe = -SdT + \tau d\theta$ where $\tau$ is the generalized ``torque'' for the bond angle $\theta$.  Our results, as shown in Fig.~\ref{fig:twodee}, display kinks in $\fe(\theta)$, indicating a discontinuity in the generalized torque.  The consequence of this discontinuity in experiments is that, at finite $T$, in order to hold in the lattice in a configuration that is not the critical state, torque has to be applied.  At zero torque, the lattice will stay at the critical state where $\partial F / \partial \theta$ jumps from negative to positive ($\theta=\pi/2$ for the example shown in Fig.~\ref{fig:twodee}).  In a simple experimental setup where only the hydrostatic pressure $p$ instead of the torque conjugate to the GH mode is controlled, our results indicate an abrupt change in pressure as the lattice passes the critical state along the GH mode.  At zero or small pressure, the lattice adopts the critical state in equilibrium.

It is worth noting that at the critical states, although the lattice has bulk FM along the direction where bonds form straight lines, the other lattice direction is still topologically polarized and exhibit asymmetric mechanical response.  For example, at $\theta=\pi/2$, the topological polarization along the horizontal direction of the lattice in Fig.~\ref{fig:twodee} is not defined, but the topological polarization in the other lattice direction is defined, leading to much greater stiffness on the top boundary than the bottom boundary of the lattice.

\section{Anharmonicity and regularization of the free energy}
\label{sec:regularization}
The analytic theory we discussed above is based on integrating out small fluctuations using a quadratic theory [keeping to $O(u^2)$ in the Hamiltonian].  Very close to the emergence of extra FM-SS pairs, fluctuations become large, which invalidates the quadratic theory.  In the case of the four-bar linkage, the extra FM is zero energy to all orders, because it represent the crossing of the zero-energy manifold itself.  Thus, very close to the bifurcation point in the critical condition of the four-bar linkage, the quadratic theory prediction of $\fe \sim k_B T \log |\theta_1(c) - \theta_1(c_0)|$ no longer applies, in the sense that the system may escape to the other mechanism at the crossing point.

There are also systems in which the mechanism softens other floppy modes, but these emergent floppy modes are only floppy to quadratic order (they are not additional mechanisms).  Examples include jammed packings of particles in which floppy modes exhibit strong anharmonicity~\cite{LiuNag2010a} and fiber networks in which floppy bending modes stiffen when fibers become collinear~\cite{BroederszMac2014,Feng2015,Feng2016}.

To understand these {finite temperature systems where other modes only soften to quadratic order along the mechanism} one needs to go beyond the quadratic theory.  
Here we first use a schematic Hamiltonian with one emergent floppy mode to discuss this effect and show that the singularity is regularized by the anharmonic terms, and then discuss the more general case of multiple coupled floppy modes using the topological kagome lattice as an example.

The schematic Hamiltonian is given by the following equation
\begin{align}\label{EQ:anH}
	H = \frac{k}{a^2} \left\lbrack (\Delta c)^2 u^2 + u^4 \right\rbrack ,
\end{align}
where $k$ is the spring constant, $a$ is some microscopic length scale (e.g., length of bonds between sites) to make dimensions right, $\Delta c = c-c_0$ with $c$ being the coordinate along the zero-energy manifold and $c_0$ being the coordinate on this manifold where the mode $u$ becomes floppy in quadratic order.  Following the similar calculation of integrating out the fluctuations $u$ as in Eq.~\eqref{eq:fe} but to quartic order in $u$, we have
\begin{align}\label{EQ:FFull}
	\fe = -k_B T \log \left\lbrack \frac{\vert \Delta c\vert }{2a} e^{Y}
	K_{1/4}\left(Y\right) \right\rbrack.
\end{align}
where we have defined the dimensionless combination $Y\equiv \frac{k(\Delta c)^4}{8a^2 k_B T}$ (which characterizes the distance to $c_0$ on the manifold normalized by thermal fluctuations), and $K_{1/4}$ is the modified Bessel function of the second kind.  A toy model {of a point mass tethered by two springs}
that exhibit this type of behavior has been discussed in Ref.~\cite{Zhang2016}.  
Using the asymptotic form of $K_{1/4}$, we find that far away from the emergence of the new floppy mode, $(k/a^2 )(\Delta c)^4\gg k_B T$, the free energy takes the form
\begin{align}\label{EQ:FFar}
	\fe \simeq k_B T \log \sqrt{\frac{k (\Delta c)^2}{\pi k_B T}}{\sim k_B T \log |\Delta c|},
\end{align}
agreeing with the quadratic theory result of Eq.~(\ref{EQ:FELOG}), with a logarithmic divergence of $\fe$ at the emergence of the new floppy mode.  

\begin{figure}
\includegraphics[width=0.49\textwidth]{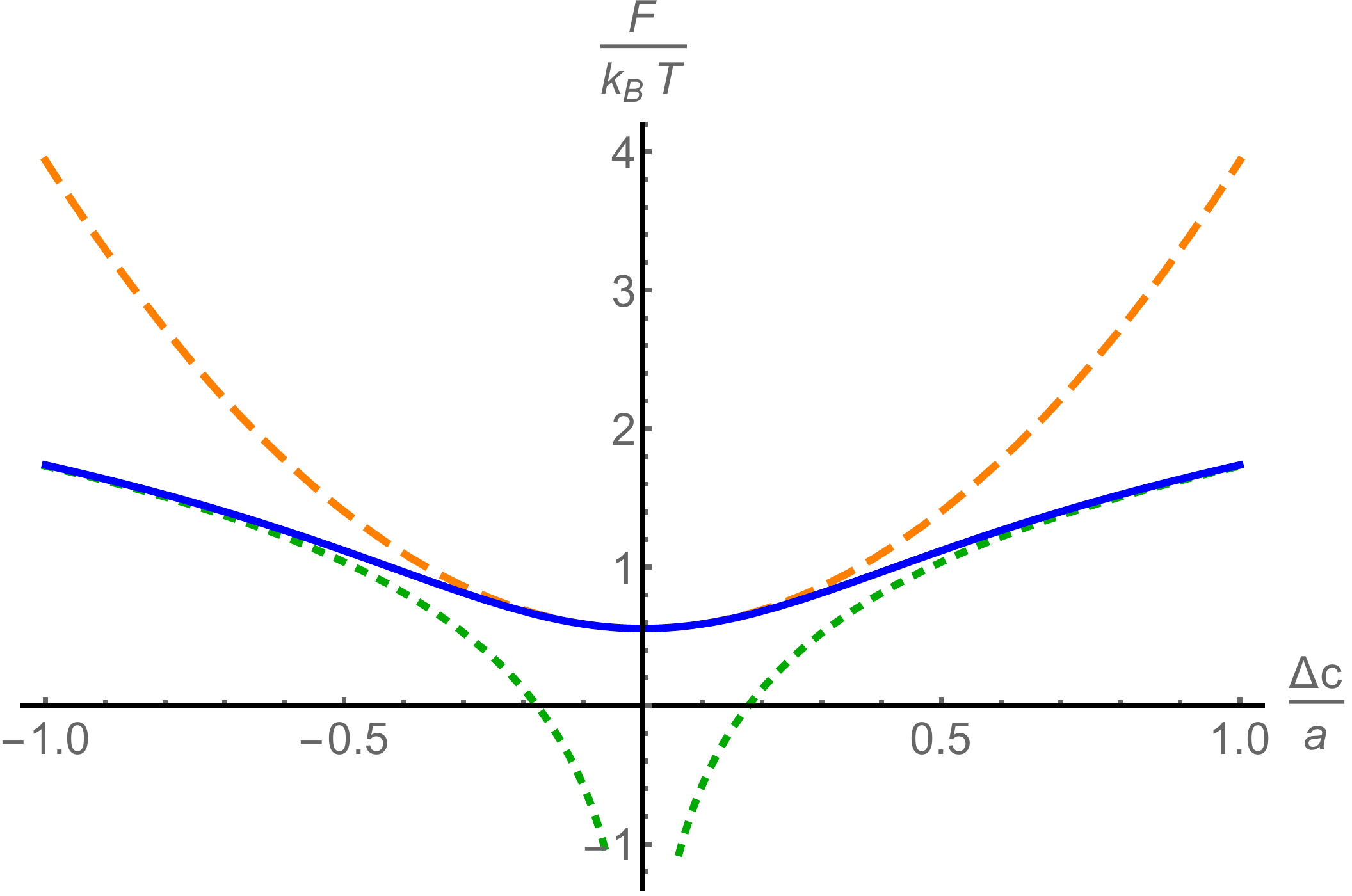}
\caption{
Free energy along a mechanism calculated from the Hamiltonian [Eq.~\eqref{EQ:anH}] which includes an anharmonic term in the fluctuations.  The exact expression Eq.~\eqref{EQ:FFull} is shown as the blue solid curve.  The asymptotic expression of the free energy far from the emergence of the new floppy mode [Eq.~\eqref{EQ:FFar}] is shown in the green dotted curve, and the asymptotic expression close to the emergence of the new floppy mode [Eq.~\eqref{EQ:FNear}] is shown in the orange dashed curve.  We take $ka^2/(k_B T)=100$ in the plot. %fig generated in finiteT.nb
}
\label{Fig:FEanhamonic}
\end{figure}

In contrast, close to the emergence of a new floppy mode, $(k/a^2 )(\Delta c)^4\ll k_B T$, the free energy takes the form
\begin{align}\label{EQ:FNear}
	\fe \simeq & k_B T \log \left\lbrack \frac{\Gamma(1/4)}{2} \left(\frac{k a^2}{k_B T}\right) ^{1/4} \right\rbrack \nonumber\\
	&+ \frac{\Gamma(3/4)}{\Gamma(1/4)}\sqrt{k_B T k a^2} \frac{(\Delta c)^2}{a^2} + O\left( \left(\frac{\Delta c}{a}\right)^4 \right),
\end{align}
which is quadratic in $\Delta c$ and the rigidity {has a fractional dependence on the temperature,  $\sqrt{T}$.}
The comparison of the free energy as a function of $\Delta c$ and its asymptotic forms is shown in Fig.~\ref{Fig:FEanhamonic}.  

The above schematic model describes the case of a single floppy mode $u$ that is softened by the mechanism $c$.  In a large system, especially in periodic lattices, in general, there can be a large number of floppy modes that become floppy at the same point along the mechanism, and this may modify the result we obtained above.  This can be illustrated by considering the 2D topological kagome lattices at finite $T$ which we introduced in Sec.~\ref{sec:twodee}.  In quadratic theory, the free energy shows singularities at the critical lattice configurations, because a whole line of phonon modes in the first Brillouin zone becomes zero frequency at these points.  It is straightforward to show that these modes are bounded by anharmonic terms, by expanding the full lattice Hamiltonian to higher order.  From the discussion above we expect this anharmonicity to soften the singularity.  Interestingly, because in this case it is a whole lines of floppy modes rather than a single one, the stiffness depend on $T$ through a fractional power that is different from $1/2$, as we show below.

For these 2D lattices, the exact integral as in Eq.~\eqref{EQ:FFull} is no longer available, because there are a large number of floppy modes that are coupled to each other at the quartic level.  Instead, following the same strategy as Ref.~\cite{Mao2015,Zhang2016}, we adopt a self-consistent field theoretic method to characterize the fluctuation correction from the anharmonic terms.  
To show this, it is convenient to introduce next-nearest-neighbor (NNN) springs, of spring constant $\kappa$.  This is simply for the purpose of facilitating the self-consistent theory, and in the end we will take $\kappa\to 0$.   Assuming that all the NNN springs are at their rest length at the critical state $\theta_c$, the Hamiltonian then exhibit a quadratic term in $\Delta\theta\equiv \theta-\theta_c$ (favoring the critical state), instead of equal to zero for any $\theta$,
\begin{align}
	H_{NNN}(\Delta\theta) = \frac{1}{2} g_1 \kappa a^2 \Delta\theta^2 + O\left((\Delta\theta)^4\right),
\end{align}
where $g_1$ is a geometric constant and $a$ is the lattice constant.  The $T=0$ stiffness of the lattice against the mechanism has only contribution from the NNN springs $d^2 H/d\Delta\theta^2 = g_1 \kappa a^2$.
We then apply the quadratic theory free energy calculation~\eqref{EQ:FtwoD} again by integrating out the phonon modes numerically.  The numerical results we obtain are well described by the asymptotic form
\begin{align}\label{EQ:AsymF}
	\fe(\Delta\theta) = &H(\Delta\theta) 
	+ k_B T \Big\lbrack g_2 + g_3 \Delta\theta  \nonumber\\
    &+ g_4\sqrt{g_5 (\kappa/k) +(\Delta\theta)^2} \Big\rbrack ,
\end{align}
where $g_2, g_3, g_4, g_5$ are geometric constants, and $k$ is the nearest-neighbor spring constant.  {Discussions of the kagome lattice phonon spectrum that leads to such asymptotic forms can be found in Ref.~\cite{SunLub2012,Mao2011a}.}
It is straightforward to see that in the limit of $\kappa\to 0$ this reduces to the quadratic theory result we had in the previous section where $F\sim k_B T (g_3 \Delta\theta+g_4|\Delta \theta|)$ [Eq.~\eqref{EQ:FE2DASYM}].     If $|g_3|<g_4$, the point $\theta_c$ is a minimum of free energy.  Otherwise it is a kink.  
The validity of this form is verified by numerical calculation of the entropy following Eq.~\eqref{EQ:FtwoD} (with the addition of NNN bonds), as shown in Fig.~\ref{Fig:kagomeF}.

\begin{figure}
\includegraphics[width=0.49\textwidth]{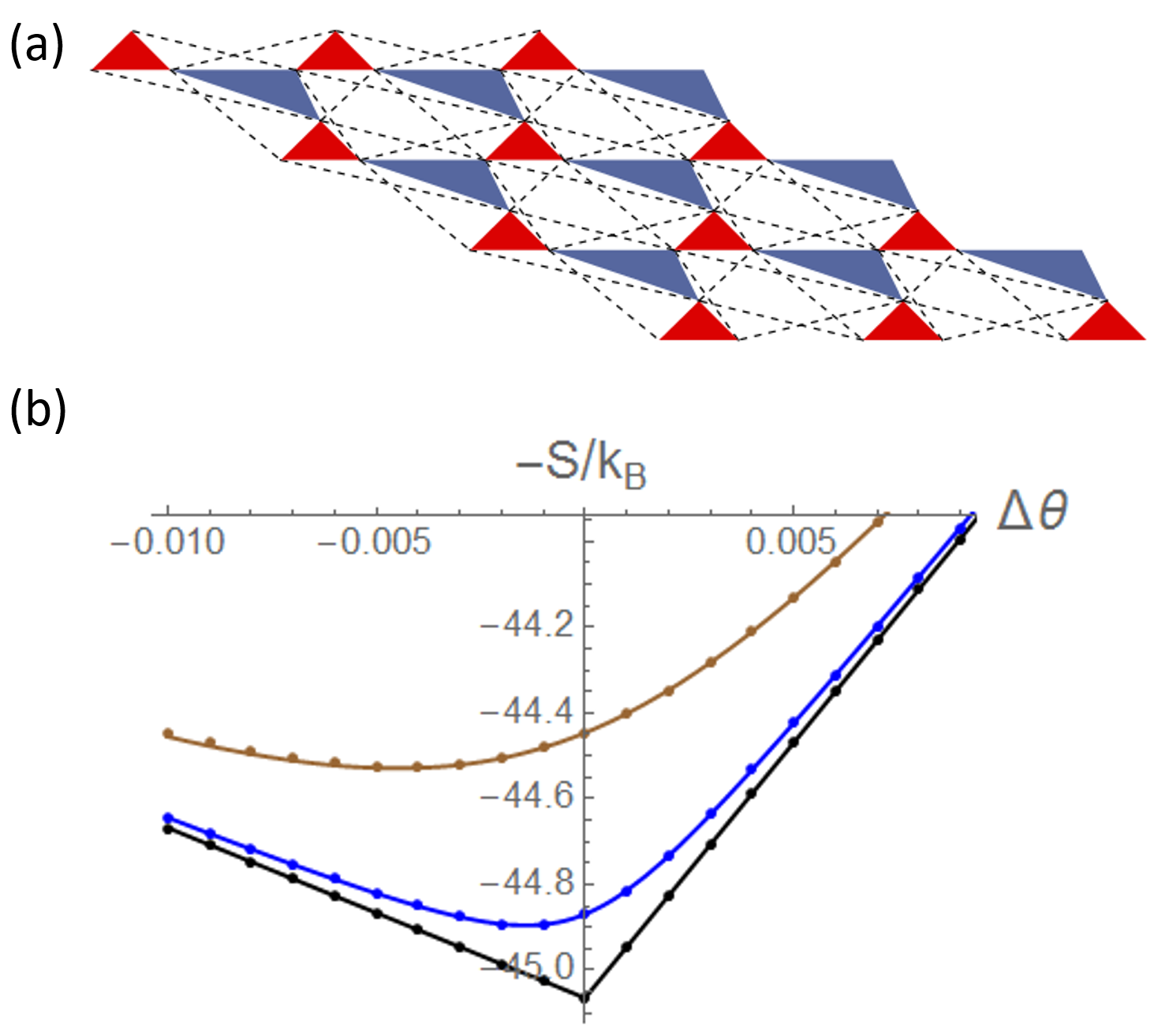}
\caption{Numerical calculation of entropic part of the topological kagome lattice free energy.  (a) The topological kagome lattice being used for calculation at the critical configuration $\theta=\theta_c$ leading to straight lines of bonds in the horizontal direction.  Unit cell geometry of the lattice is chosen such as the red (pointing up) triangles have side lengths $(0.7071,1,0.7)$ and the blue (pointing down) triangles have side length $(0.7142,1.5780,2)$.  NNN bonds are shown in black dashed lines.  (b) Entropic part of the lattice free energy per unit cell normalized by $k_B$  as a function of $\Delta \theta$ (points).  From bottom to top the points correspond to $\kappa=0$ (black), $\kappa/k=10^{-6}$ (blue), and $\kappa/k=10^{-5}$ (brown).  The curves show the asymptotic form as in Eq.~\ref{EQ:AsymF} with $g_3=39.66$, $g_4=79.33$, $g_5=6.075$, where $g_3$ and $g_4$ are determined using the $\kappa=0$ data, $g_5$ is determined using $\kappa/k=10^{-6}$ data.  Using these values we have good agreement with data of $\kappa/k=10^{-5}$.
%generated in topological_kagome_T2_paper1.nb
}
\label{Fig:kagomeF}
\end{figure}

The self-consistency of the field theory arises from
 the fact that the mechanism $\theta$ (the GH mode which is a homogeneous deformation) and the floppy phonon modes (which are of finite wavelength) are of the \emph{same origin}.  As shown in Fig.~\ref{fig:twodee}, they both involve rotating triangles around the hinged sites in the same pattern, with the only difference being that the GH mode is a homogeneous rotating mode of all triangles throughout the lattice, whereas the floppy phonon modes involve spatially varying rotation.   
This can also be seen from the fact that both the mechanism and the floppy phonon modes have rigidity proportional to $\kappa$ (at $T=0$ and $\theta=\theta_c$), because they both deform the NNN springs only.  
The stiffness of the NNN spring constant is renormalized at finite $T$ by thermal fluctuations, and this effect can be extracted by taking derivatives of $F$ with respect to $\theta$, leading to the renormalized NNN spring constant
\begin{align}\label{EQ:ReR}
	r = \frac{1}{g_1 a^2} \frac{d^2 \fe}{d\theta^2}\Big\vert_{\theta_c} = \kappa 
+ \frac{g_4}{g_1 \sqrt{g_5}a^2} \frac{k_B T}{\sqrt{\kappa/k}} .
\end{align}

As we discussed above, because the mechanism $\theta$ and the floppy modes are of the same origin, we can make the self-consistent approximation, and replace $\kappa$ on the right hand side of Eq.~\eqref{EQ:ReR}.  Taking $\kappa\to 0$ we have the regularized stiffness of the mechanism
\begin{align}
	\frac{d^2 \fe}{d\theta^2}\Big\vert_{\theta_c} = \frac{g_4}{\sqrt{g_5}}  (ka^2)^{1/3} (k_B T)^{2/3},
\end{align}
carrying an anomalous entropic rigidity exponent of $2/3$, differing from the $1/2$ exponent we obtained for the single floppy mode case.  The $2/3$ exponent agrees with our previous results on non-topological square lattice and kagome lattice~\cite{Mao2015,Bedi2018}.
It is worth pointing out that such fractional dependence on $T$ of the rigidity of a mechanism (e.g., shear deformation) has been observed in various systems in which the mechanism couples to other floppy modes in the system~\cite{DennisonMac2013,Mao2015,Zhang2016,Degiuli2015}.

To summarize, in presence of anharmonic terms, fluctuations of floppy modes are bounded, and their correction to the rigidity of the mechanism shows quadratic form near the point where the mode becomes soft, {$\fe\sim K_T (c-c_0)^2$}, as opposed to the logarithmic divergence as observed in the quadratic theory.  The coefficient $K_T$ of the quadratic free energy, which is the thermal stiffness of the lattice against deformation around the free energy minimum, is a function of $T$ with nontrivial exponent, {$K_T\sim T^{\beta}$}, depending on the floppy mode structure of the system.

\section{Implementation}
\label{sec:implementation}

\begin{figure*}
\centering
\subfigure{\includegraphics[width=.45\textwidth]{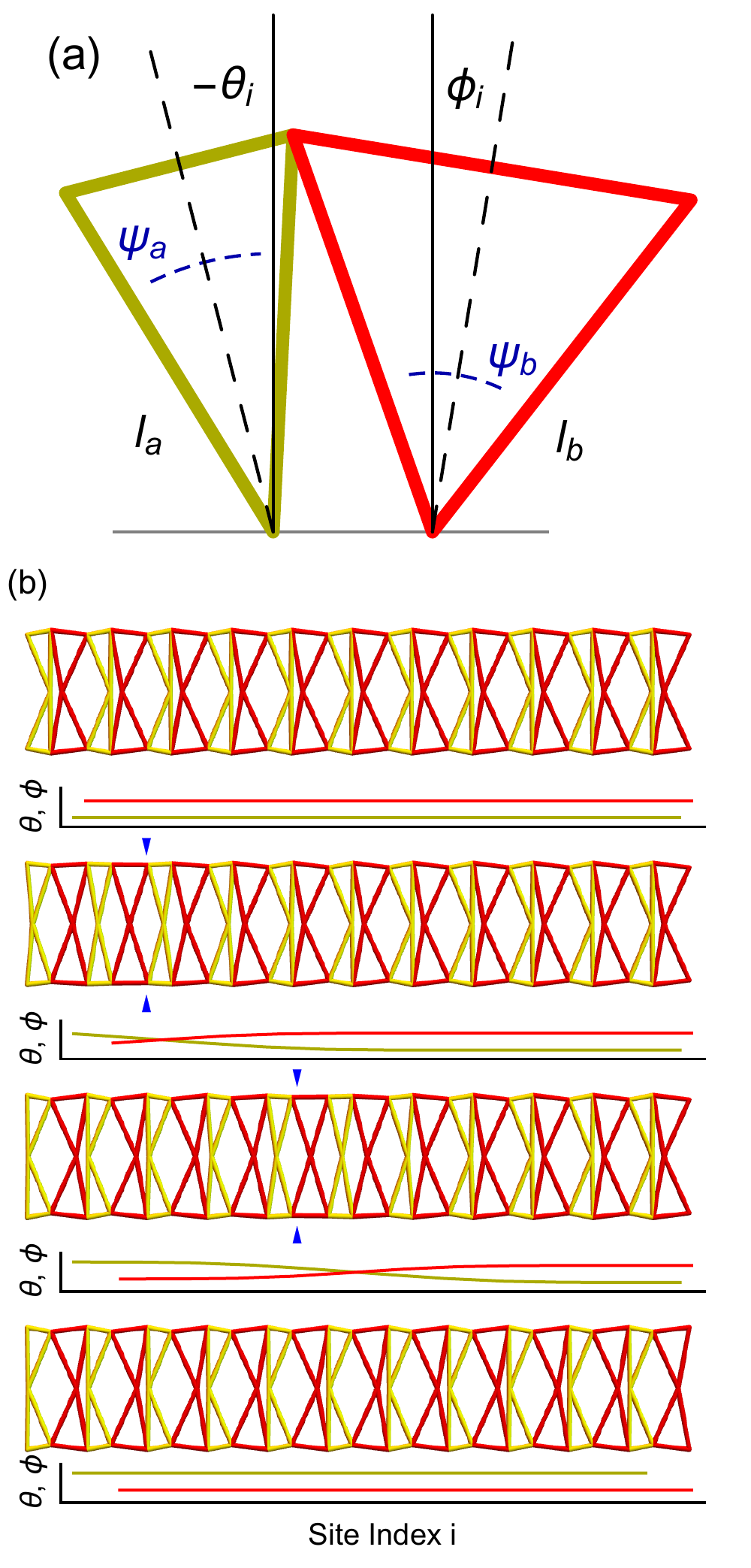}}
\subfigure{\includegraphics[width=.53\textwidth]{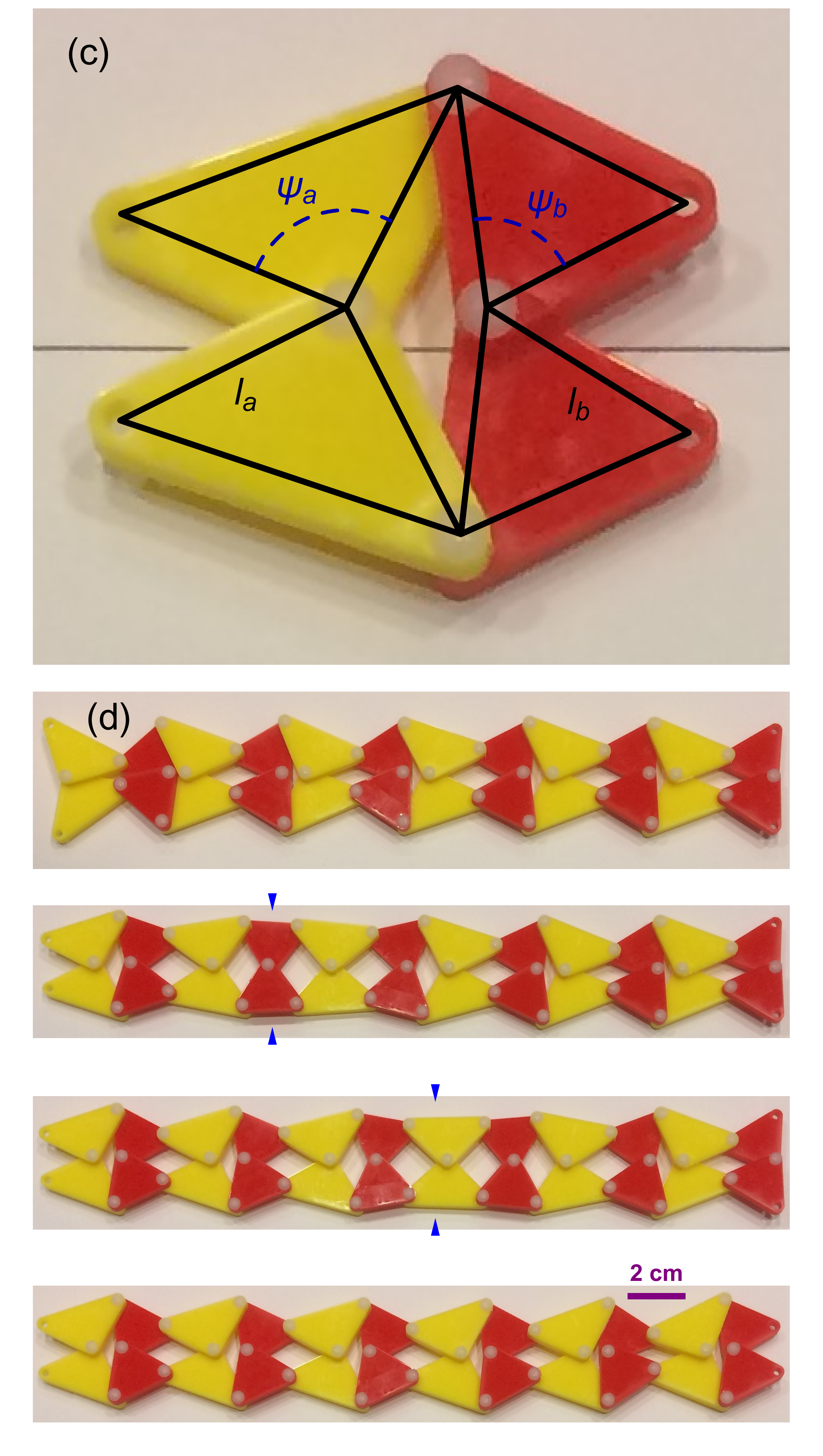}}
\caption{
(a) The components of a triangle chain system consist of rigid isosceles triangles free to rotate at joints, assuming angles $\tangone_i,\tangtwo_i$ to the vertical. The two types of triangles are characterized by side lengths $\lena, \lenb$ and angular widths $\anga, \angb$.
(b) A single triangle chain is shown in four configurations, with a soliton indicated by blue triangles moving from left to right across the system, transforming it from a configuration in which the yellow triangles point to the left to one in which they point to the right. Plots under the images describe the orientations of the triangles.
(c) A segment of a centimeter-scale prototype consisting of rigid triangles joined at free hinges.
(d) A soliton indicated by blue triangles moves across the prototype under manual manipulation.
}
\label{fig:trichain}
\end{figure*}

\begin{figure}
\includegraphics[width=0.49\textwidth]{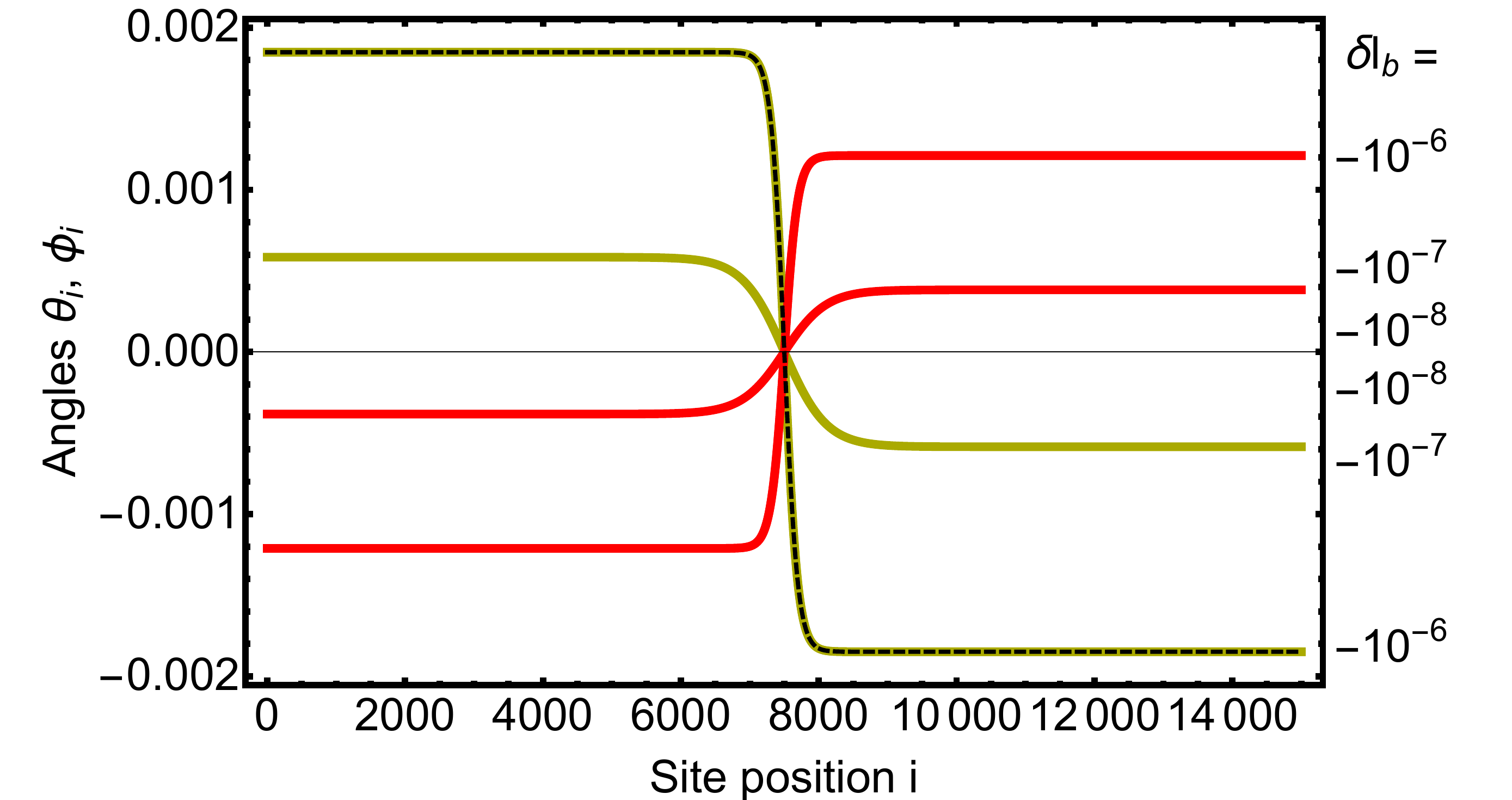}
\caption{The angles $\tangone_i,\tangtwo_i$ of sites $i$ in configurations with solitons. These chains differ in
$\delta \lenb$, the difference of the edge length from that which would permit uniformly vertical triangles ($\tangone_i=\tangtwo_i=0$), with the soliton becoming flatter and wider as this limit is approached.
 The dashed line shows good agreement with the hyperbolic tangent profile of a $\phi^4$ chain. }
\label{fig:multwidths}
\end{figure}

In previous sections, we envision mechanical devices consisting of regular sections and driven by thermal fluctuations. Macroscopic degrees of freedom are too large to be driven by thermal fluctuations, but nevertheless can acquire ``effective temperatures'' either from external forces such as vibrations~\cite{jaeger1992physics} or by internal activity~\cite{loi2008effective}. Such forces could be used to drive the structures we describe into precise configurations such as expanded and collapsed states. However, it is at the microscopic scale at which thermal mechanisms become ubiquitous and unavoidable.

Microscopic mechanical structures consisting of repeated rigid units are readily achieved via \emph{DNA origami}~\cite{rothemund2006folding}. DNA origami uses oligonucleotides to fold DNA strands into desired shapes, with self-assembly permitting large structures, extending to hundreds of nanometers. The combination of stiff double-stranded DNA and flexible single-stranded DNA can create structures with flexible hinges that permit large mechanisms~\cite{Marras2015}. These mechanisms can be reversibly actuated by the reversible addition of further DNA strands that apply forces at the joints. Without such additional strands, four-bar linkages were observed to fluctuate around their zero-energy mechanism consistent with the picture described in Sec.~\ref{sec:fourbar}. However, these fluctuations, of $\approx10^\circ$, occur in a linkage of beams connected loosely at joints and permitted to move in three dimensions, precluding quantitative comparison with the entropic effects of our simpler model. Other studies also observe thermal fluctuations in the angles of flexible DNA structures on the order of $1^\circ-10^\circ$~\cite{dietz2009folding, castro2011primer}.

To go from the four-bar linkage to more complicated structures capable of conveying force, motion and information over extensive distances requires a DNA origami chain. To pass beyond the rotor chain of Fig.~\ref{fig:onedeechain} to a free-standing DNA origami structure, we 
now present the \emph{triangle chain}. This structure has many degrees of freedom but only one floppy mode, so that its zero-energy manifold is one-dimensional but embedded in a large-dimensional space.
It consists of two types of pairs of isosceles triangles, joined at their hinges and constrained to move in two dimensions. The triangles have side lengths $\lena, \lenb$ and angular widths $\anga, \angb$, leading to the nonlinear zero-energy transfer relations between the angles of the two types of triangles $\tangone_i,\tangtwo_i$ from the vertical:

\begin{align}
\label{eq:trichain}
\lena \cos \left( \tangone_i +\anga/2\right) &= \lenb \cos \left( \tangtwo_i -\angb/2\right), \\
\lenb \cos \left( \tangtwo_i +\angb/2\right) &= \lena \cos \left( \tangone_{i+1} -\anga/2\right).
\end{align}

\noindent As with the rotor chain, this system has a kink mechanism of finite width that flips the chain between two uniform configurations related by a mirror symmetry. Unlike the rotor chain, this system is translationally invariant rather than fixed to a substrate and operates in two dimensions without self-intersections, making it realizable via DNA origami. Applying the design principles discussed in Sec.~\ref{sec:chain}, one can choose geometric parameters to ensure that the soliton mechanism operates smoothly even while undergoing thermal fluctuations.

To demonstrate the mechanics of the triangle chain, we have created a prototype composed of laser-cut centimeter-scale hard PMMA triangles as shown in Fig.~\ref{fig:onedeechain} (c,d). Manual manipulation thereof is demonstrated in a Supplementary Video. As shown in Fig.~\ref{fig:multwidths}, altering the shapes of the triangles can generate solitons whose widths substantially exceed the lattice spacing of the chain.

\section{Conclusion}

We have shown that thermal fluctuations can significantly modify the behavior of mechanical frames near the point of mechanical stability. Conformational changes alter the phonon spectrum, generating an entropic force that acts on the system as it passes through energetically degenerate states. This effect tends to align bonds to render the constraints as redundant as possible, permitting the largest fluctuations in the remaining modes. However, when additional zero modes appear, fluctuations become nonlinear, indicating additional conformational changes due to a change in the topology of the zero-energy manifold. At sufficiently high temperature, this change in topology can occur even off the critical point due to thermal tunneling through finite-energy states.

This general behavior manifests itself in sharply different ways for different systems. For one-dimensional chains, a discrete (crystal) translational symmetry prevents the free energy from increasing as the soliton moves through the bulk. Instead, the modifications to the phonon spectrum are periodic and exponentially small in the soliton width. Provided the soliton is narrow enough or the materials stiff enough to prevent thermal tunneling into alternate states, the soliton proceeds largely unimpeded by thermal effects.

In contrast, thermal fluctuations substantially modify the mechanics of two-dimensional Maxwell lattices, which have a large number of zero modes. These fluctuations tend to drive the lattices to a critical state between two different topological polarizations, one in which the zero modes that would typically lie on the edge are instead in the bulk. Thus, the fluctuations order the initially floppy lattice and grant it an entropic rigidity.

The essential physics considered here is that of a thermal system with some permanent structure but also one or more instabilities. This extends beyond the simple mechanical frames presented here. It would be interesting to consider mechanical systems with other degrees of freedom such as origami/kirigami~\cite{Chen2015b}, orientational degrees of freedom~\cite{Paulose2015}, three-dimensional structure~\cite{Stenull2016}, etc. Non-mechanical systems, such as spin antiferromagnets~\cite{Lawler2013}, also have similar physics.

To sum up, our work provides a geometrical design blueprint to control thermal fluctuations in miniaturized mechanical structures that are deployed or reconfigured using folding mechanisms. We show (i) how to suppress thermal fluctuations so that useful kinematic mechanisms 
remain unobstructed or, conversely, (ii) how to exploit them to thermally lock reconfigurable nano-mechanical devices into a desired structure. We envision application of these design principles to the engineering of nano-mechanical devices based on DNA origami or activated mechanisms that exploit molecular robots.  

\emph{Acknowledgments--} The authors acknowledge helpful conversations with Deshpreet Bedi. D.Z.R. was supported by the  ICAM postdoctoral fellowship, the Bethe/KIC Fellowship and the National Science Foundation through Grant No. NSF DMR-1308089. X.M. acknowledges support from the National Science Foundation under grants numbers NSF-DMR-1609051 and NSF-EFMA-1741618. V.V. was primarily supported by the University of Chicago Materials Research Science and Engineering Center, which is funded by the National Science Foundation under award number DMR-1420709.

\bibliography{toptemp}

\appendix
\section{Free energy of 1D chains}\label{APP:FE}

Consider a general linear coupling between neighboring degrees of freedom $\{\disp_j\}$ and spring extensions $\{\ext_j\}$:

\begin{align}
\label{eq:linrel}
\ext_j = a_j \disp_{j} - b_j \disp_{j+1}.
\end{align}

\noindent As discussed in the main text, the free energy is proportionate to the sum of the logarithms of the normal frequencies of the system, which may be found from the determinant of the dynamical matrix. We could evaluate this by freezing either the rightmost or the leftmost site, so that the rigidity matrix is upper/lower diagonal and the determinant of the dynamical matrix is easily evaluated to be $\prod_j{a_j}$ or $\prod_j{b_j}$. Since these two expressions are generally unrelated, this indicates an extraordinary degree of dependence upon the boundary conditions of the system. Instead of these fixed boundary conditions, we choose open ones such that we have $\numsites$ sites and $\numsites-1$ bonds. This means we will necessarily have a zero mode, and to evaluate the finite part of the free energy in light of it, we find the \emph{pseudodeterminant}, the product of the nonzero eigenvalues, obtained via

\begin{align}
D_\numsites \equiv  \lim_{\alpha \rightarrow 0}
\textrm{det}\left(
\dmat(\{a_j\},\{b_j\})+\alpha \mathbf{I}\right).
\label{eq:pseudod}
\end{align}

The dynamical matrix, obtained via the relation Eq.~(\ref{eq:linrel}), and related matrix which we define now are

\begin{align}
\dmat_\numsites = 
\begin{bmatrix}
    a_1^2       & a_1 b_1 & 0 & \ldots & 0 \\
    a_1 b_1       & a_2^1+b_1^2 & \ddots & \dots & 0 \\
    \vdots 	& \ddots & \ddots & \ddots & 0 \\
    0 & 0 & a_{\numsites-1} b_{\numsites-1} & a_\numsites^2 + b_{\numsites-1}^2 & a_{\numsites} b_{\numsites} \\
		0 & 0 &   0 & a_{\numsites} b_{\numsites} & b_{\numsites}^2 
\end{bmatrix} 
\\
\emat_\numsites \equiv
\begin{bmatrix}
    a_1^2       & a_1 b_1 & 0 & \ldots \\
    a_1 b_1       & a_2^1+b_1^2 & \ddots & \dots  \\
    \vdots 	& \ddots & \ddots & \ddots  \\
    0 & 0 & a_{\numsites-1} b_{\numsites-1} & a_\numsites^2 + b_{\numsites-1}^2 
\end{bmatrix} .
\end{align}

\noindent $\emat_\numsites$, lacking the final column and final row of $\dmat_\numsites$, corresponds physically to a system in which the final site is frozen and not allowed to move. 
Even without such a physical interpretation, we may use the method of minors to obtain $\emat_\numsites$'s pseudo-determinant and relate it to the desired $D_\numsites$. Expanding Eq.~(\ref{eq:pseudod}) along the bottom row results in the recursion relationship

\begin{align}
\label{eq:pdlinearterm}
D_\numsites = \left(b_\numsites^2 + \alpha\right) M_\numsites - a_\numsites^2 b_\numsites^2 \en_{\numsites-1}=
\\ \nonumber
b_\numsites^2\left(M_\numsites^0 - a_\numsites^2 M_{\numsites-1}^0\right) +
\alpha \left[
M_\numsites^0 + b_\numsites^2 
\left(
M_\numsites' - a_\numsites^2 M_{\numsites-1}'\right)
\right],
\end{align}

\noindent where we define

\begin{align}
\textrm{det}\left(
\emat(\{a_j\},\{b_j\})+\alpha \mathbf{I}\right)\equiv M_\numsites^0 + \alpha M_\numsites' + O(\alpha^2).
\end{align}

\noindent Because we have a zero mode, the $O(\alpha^0)$ portion of this determinant must vanish, which requires, given $M_1 = a_1^2$, that $M_\numsites^0 = \left(a_1^2 a_2^2 \ldots a_\numsites^2\right)$. By again evaluating determinants via the blocking method, we obtain the additional relationship

\begin{align}
\label{eq:mmatrecur}
M_\numsites' - a_\numsites^2 M_{\numsites-1}' = 
\\ \nonumber
M_{\numsites-1}^0 + b_{\numsites-1}^2 \left(
M_{\numsites-1}' - a_\numsites^2 M_{\numsites-2}'\right) + O(\alpha^1)
.
\end{align}

\noindent We may now repeatedly apply the recursion relationship in Eq.~(\ref{eq:mmatrecur}) into the leading, $O(\alpha^1)$ term in Eq.~(\ref{eq:pdlinearterm}) to obtain our final expression:

\begin{align}
D_\numsites = \sum_{k=0}^{\numsites} 
\left(
\prod_{j=1}^{j=k} a_j^2 \prod_{j=k+1}^{j=\numsites} b_j^2
\right)
 = \\ \nonumber
(a_1^2 a_2^2\ldots \ldots a_{\numsites-1}^2 a_\numsites^2) + 
(a_1^2 a_2^2\ldots \ldots a_{\numsites-1}^2 b_\numsites^2) + \\ \nonumber
(a_1^2 a_2^2\ldots \ldots b_{\numsites-1}^2 b_\numsites^2) +
 \ldots +
(b_1^2 b_2^2\ldots \ldots b_{\numsites-1}^2 b_\numsites^2)
\end{align}.

\noindent Ignoring lattice effects, this results in the continuum free energy

\begin{align}
\fe = \frac{\temp}{2} \int_{x_s}^{x_f} d x \exp \left( \int_{x_s}^x \log a(x') + \int_x^{x_f} \log b(x') \right).
\end{align}

\section{Exponential free energy barriers for smooth kinks}\label{APP:EXPO}

In the main text, we consider a one-dimensional chain of mechanical bonds that, at zero temperature, contains a soliton-like kink. At finite temperature, thermal fluctuations grant a finite free energy.
We are then presented with an expression for the free energy of the form

\begin{align}
\mathcal{F}(x_c) = \frac{\temp}{\spacing}
\int_{-\infty}^{\infty} d x \,
\frac{\cos(2 \pi x/\spacing)}{f((x-x_c)/\width)} 
\\
=  \temp \frac{ \width}{\spacing} 
\cos \left( \frac{2 \pi x_c}{\spacing}\right)
 \int_{-\infty}^{\infty}
 d y \, \frac{\cos\left[(2 \pi \width/\spacing)y\right]}{f(y)},
\end{align}

\noindent where $\spacing$ is the lattice spacing, $\width$ is the width of the soliton and $x_c$ is the position of its center.  

Equation~(\ref{eq:chainfe}) is an example of this free energy form when the soliton profile is described by Eq.~(\ref{eq:soliton}).  In this Appendix, we
consider the general case in which the soliton does not possess any zero modes, so that $f(\cdot)$ is finite at any point along the line. Furthermore, we restrict ourselves to analytic functions with only isolated zeroes in the complex plane and whose magnitude falls off fast enough to permit the use of contour integrals. We also assume, primarily for convenience that $f(-z) = f(z)$. This permits replacing the cosine with a complex exponential and using a single contour.
 Thence, we may evaluate via the Residue Theorem to obtain

\begin{align}
\mathcal{F}(x_c) = 
2 \pi i \temp\frac{ \width}{\spacing} \cos \left( 2 \pi x_c /a\right)
\sum_j \textrm{Res}_j,
\end{align}

\noindent where $\textrm{Res}_j$ are the residues of the expression $\exp(2 \pi i (\width/\spacing)y/f(y)$ in the complex upper half-plane.
From this form, it is apparent that the free energy costs associated with moving the soliton are greatest when the motion of the soliton nearly admits zero modes into the lattice, so that some $z_j$ have small imaginary parts. Otherwise, we see that the free energy cost is exponentially small in the soliton width, by a factor $\exp \left[2 \pi (\width/\spacing)\textrm{Im}(y^*)\right]$, where $y^*$ is the location of the residue with smallest positive imaginary part. Hence, solitons travel freely when they take place over multiple lattice spacings and when they don't couple too closely to additional zero modes. We may repeat this procedure for more oscillatory lattice terms, obtaining exponentially smaller corrections.

\end{document}